\documentclass[12pt,a4paper]{article}
\usepackage{graphicx}
\usepackage{times}
\title{\bf Properties of Narrow line Seyfert 1 galaxies}
\author{Suvendu Rakshit$^1$\thanks{suvenduat@gmail.com}, C. S. Stalin$^1$, Hum Chand$^2$ and Xue-Guang Zhang$^3$\\
\vspace{1cm}\\
\normalsize $^1$ Indian Institute of Astrophysics,  Block II, Koramangala, Bangalore-560034, India \\ 
\normalsize $^2$ Aryabhatta Research Institute of Observational Sciences (ARIES), 263002, Nainital, India\\
\normalsize $^3$ Institute of Astronomy and Space Science, Sun Yat-Sen University, No. 135, Xingang Xi Road, \\
\normalsize Guangzhou, 510275, P. R. China}
\date{\mbox{}}
\begin{document}
\maketitle
\pagestyle{empty}
%
%
\def\bull{\vrule height .9ex width .8ex depth -.1ex}
\makeatletter
\def\ps@plain{\let\@mkboth\gobbletwo
\def\@oddhead{}\def\@oddfoot{\hfil\scriptsize\bull\quad
"First Belgo-Indian Network for Astronomy \& astrophysics (BINA) workshop'', held in Nainital (India), 15-18 November 2016 \quad\bull}%
\def\@evenhead{}\let\@evenfoot\@oddfoot}
\makeatother
%
%
\def\beginrefer{\section*{References}%
\begin{quotation}\mbox{}\par}
\def\refer#1\par{{\setlength{\parindent}{-\leftmargin}\indent#1\par}}
\def\endrefer{\end{quotation}}
%
%
{\noindent\small{\bf Abstract:}
Narrow line Seyfert 1 (NLSy1) galaxies constitute a class of active galactic nuclei characterized by 
the full width at half maximum (FWHM) of the H$\beta$ broad emission line 
$<2000$ km s$^{-1}$ and the flux ratio of [O III] to H$\beta$ $<3$. Their properties 
are not well understood since only a few NLSy1 galaxies were known earlier. 
We have studied various properties of NLSy1  galaxies using an enlarged
sample and compared them with the conventional broad-line Seyfert 1 (BLSy1) 
galaxies. Both the sample of sources have $z\le 0.8$ and their 
optical spectra from SDSS-DR12 that are used to derive various physical
parameters have a median signal to noise 
(S/N) ratio $>$10 pixel$^{-1}$. Strong correlations between the H$\beta$ and 
H$\alpha$ emission lines are found both in the FWHM and flux. The nuclear continuum luminosity is found to be 
strongly correlated with the luminosity of H$\beta$, H$\alpha$ and [O III] 
emission lines. The black hole mass in NLSy1 galaxies is lower compared to 
their broad line counterparts. Compared to BLSy1 galaxies, NLSy1 galaxies
have a stronger Fe II emission and a higher Eddington ratio that place them in
the extreme upper right corner of the $R_{4570}-\xi_{\mathrm{Edd}}$ 
diagram. The distribution of the radio-loudness parameter (R) in NLSy1 
galaxies drops rapidly at $R>10$ compared to the BLSy1 galaxies that have 
powerful radio jets. The soft X-ray photon index in NLSy1  galaxies is on 
average higher ($2.9\pm0.9$) than BLSy1 galaxies ($2.4\pm0.8$). It is 
anti-correlated with the H$\beta$ width but correlated with the Fe II 
strength. NLSy1 galaxies on average have a lower amplitude of optical 
variability compared to their broad lines counterparts. These results
suggest Eddington ratio as the main parameter that drives optical 
variability in these sources.}

%
%
\section{Introduction}
Narrow-line Seyfert 1 (NLSy1) galaxies form a peculiar class of active 
galactic nuclei (AGNs) that are classified based on their optical emission line
properties. They have their full width at half maximum (FWHM) of the 
H$\beta$ broad emission line $<2000$ km s$^{-1}$, weak [O III] emission 
lines with a flux ratio of [O III] to H$\beta <3$ and strong Fe II emission 
relative to H$\beta$ (Osterbrock \& Pogge 1985). They show high amplitude 
rapid X-ray variability and a strong soft X-ray excess (Boller et al. 1996; 
Leighly 1999). They harbor low mass black holes ($M_{\mathrm{BH}}$, $10^6 - 10^8 \, M_{\odot}$) 
and have high Eddington ratio (Zhou et al. 2006; Xu et al. 2012;
 Rakshit et al. 2017a). However, some recent studies suggest that NLSy1 
galaxies have $M_{\mathrm{BH}}$ similar to BLSy1 galaxies and the 
main difference between them is due to geometrical effects (see Baldi et 
al. 2016 and references therein). 

Although the original definition of NLSy1 galaxies is based on a sharp cutoff 
in the broad emission line width at 2000 km s$^{-1}$, there seems to be no such 
dividing line in the line width distribution of Balmer lines which 
smoothly merges with the line width distribution of BLSy1 galaxies 
(Turner et al. 1999). Moreover, some BLSy1 galaxies with H$\beta$ FWHM $>2000$ 
km s$^{-1}$ also exhibit strong Fe II emission and soft X-ray variability 
(see Grupe et al., 1999). Therefore, studies of various properties of NLSy1 
galaxies over a large sample is necessary to understand the similarities
and differences of their properties with the conventional BLSy1 galaxies.

In this paper, we present our results on a comparative study of the emission 
line properties of a sample of NLSy1 and 
BLSy1 galaxies using a large sample of ``QSOs'' having a redshift $z<0.8$ 
from SDSS-DR 12 and discuss their radio, X-ray and optical variability 
properties. This paper is structured as follows. The spectral analysis 
is described in Section 2 followed by the optical emission line properties 
in Section 3. The radio, X-ray and optical variability properties of NLSy1 
and BLSy1 galaxies are discussed in Sections 4, 5 and 6, respectively. The
conclusions are given in Section 7. 

\section{Spectral analysis}
Recently, Rakshit et al. (2017a) have compiled a new catalog of NLSy1 galaxies 
consisting of 11,101 objects (a five fold increase from the existing 
catalog of NLSy1 galaxies, Zhou et al. 2006) after carefully and systematically fitting their 
continuum and emission line profiles. In addition to NLSy1 galaxies, the 
spectral fitting process allowed the authors to also obtain a sample of 
37,441 BLSy1 galaxies. All the continuum and emission line information obtained from that fitting is used here.    

\begin{figure}[h]
\centering
\includegraphics[width=14cm, height=6cm]{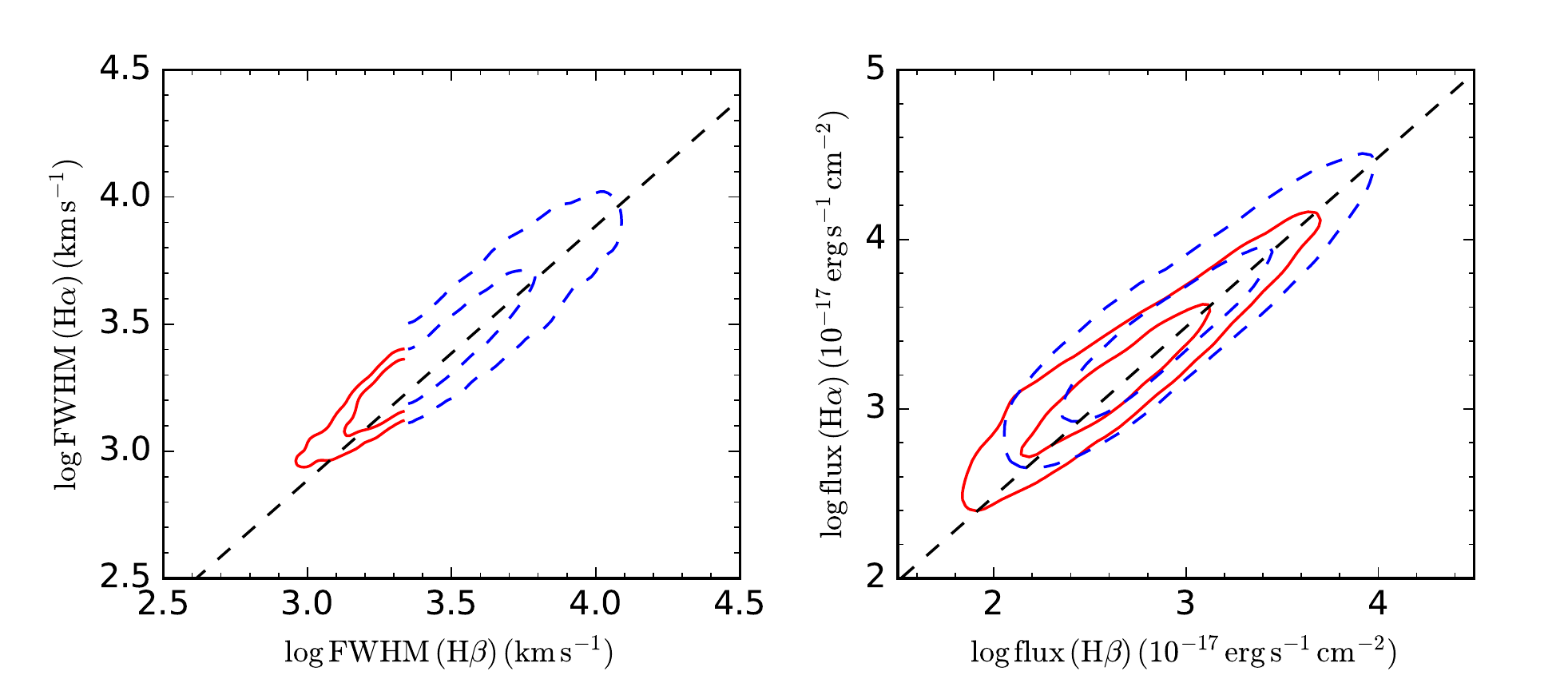}
\caption{Left: Plot of FWHM(H$\alpha$) against FWHM(H$\beta$) for NLSy1 
galaxies (solid red contours) and BLSy1 galaxies (dashed blue contours). 
Right: Flux (H$\alpha$) plotted against flux (H$\beta$). The dashed black line 
is the least squares fit to the data. The 68 percentile (inner) and 95 percentile (outer) density contours are shown in each panel.}\label{fig_1}
\end{figure}

The detailed spectral fitting procedure is described in 
Rakshit et al. (2017a). However, we summarize them briefly below. SDSS-DR12 
spectra were obtained for all the objects classified as ``QSOs'' 
by the SDSS-DR12 pipeline and having $z<0.8$. This gave 114,806 objects out of 
which many objects have a poor S/N ratio. Objects with a median S/N $>2 \, 
\mathrm{pixel}^{-1}$ were considered for further analysis which resulted in 
68,859 sources. Systematic spectral fitting was done on these sources 
via a two step  process:
\begin{enumerate}
\item {\bf Host galaxy decomposition:} The optical spectrum of low redshift AGNs will have significant contribution from their host galaxy star light, and
therefore to properly analyse their emission line properties, one needs to
remove the star light contribution. The continuum of each of the SDSS spectrum 
was therefore fitted using two components; (i) a power law AGN component and 
(ii) a stellar contribution from the host galaxy. AGN emission lines were masked 
except the Fe II multiplets during the fitting. Simple Stellar 
Population (SSP) templates from Bruzual \& Charlot (2003) were used to model 
the host contribution in the SDSS spectra. A total of 39 templates having 
ages 5  $-$ 12Gyr and solar metallicities $Z=0.008$, 0.05 and 0.02 were 
used. The AGN continuum was modeled as a power law. The continuum fitting was 
performed using the Levenberg-Marquardt least-squares minimization implemented
in the IDL  routine ``mpfit'' fitting package allowing us to decompose the 
host contribution from SDSS spectra. We then removed the stellar contribution 
leaving only the AGN contribution. 
    
\item {\bf Emission line fitting:} After removing the stellar contribution from the SDSS spectra, several lines in the H$\beta$ and H$\alpha$ regions were fitted along with a Fe II template from Kovacevic et al. (2010) and the local AGN continuum with a power law (around the H$\beta$ and H$\alpha$ regions). The lines fitted in the H$\beta$ region are broad and narrow components of H$\beta$, He II$\lambda4687\AA$ and [O III]$\lambda4959,5007\AA$ doublet while those in the H$\alpha$ region are broad and narrow H$\alpha$, [O I]$\lambda6300, 6363 \AA$, narrow [N II]$\lambda6548,6583 \AA$ doublet and the narrow [S II]$\lambda6716,6731 \AA$. Both the H$\alpha$ and H$\beta$ regions were fitted simultaneously for objects having  $z < 0.3629$. 
\end{enumerate}      

Though our initial sample was the result of fitting all spectra with a median S/N 
$>$ 2 pixel$^{-1}$, in this work, we consider only those objects having
a median S/N $>$ 10 pixel$^{-1}$ and the Fe II strength ($R_{4570}$; ratio of
Fe II line flux in the wavelength range 4434 $-$ 4684 
\AA \, to H$\beta$ flux) $>$ 0.01. This S/N cut is imposed so as to 
have unambiguous estimate of the emission line parameters.
This yielded 4070 NLSy1 and 14,314 BLSy1 galaxies, which form the 
sample analysed in this work.

\section{Emission line properties}
To study the emission line properties we have plotted in Figure \ref{fig_1} the FWHM(H$\alpha$) against FWHM(H$\beta$) on the left panel both for NLSy1 (red contours) and BLSy1 galaxies (blue dashed contours). The 68 percentile (inner) and 95 percentile (outer) density contours are shown. A least squares fit of the full sample yields $\mathrm{FWHM(H\alpha)=(0.768 \pm 0.004) \times FWHM(H\beta)}$ suggesting a strong correlation between the two parameters. On the right panel, we have plotted the flux(H$\alpha$) against the flux(H$\beta$). A least squares fit gives the relation: $\mathrm{flux(H\alpha)=(3.09 \pm 0.01) \times flux(H\beta)}$. These relations are consistent with the finding of Zhou et al. (2006) and others.

\begin{figure}[h]
\centering
\includegraphics[width=16cm, height=5cm]{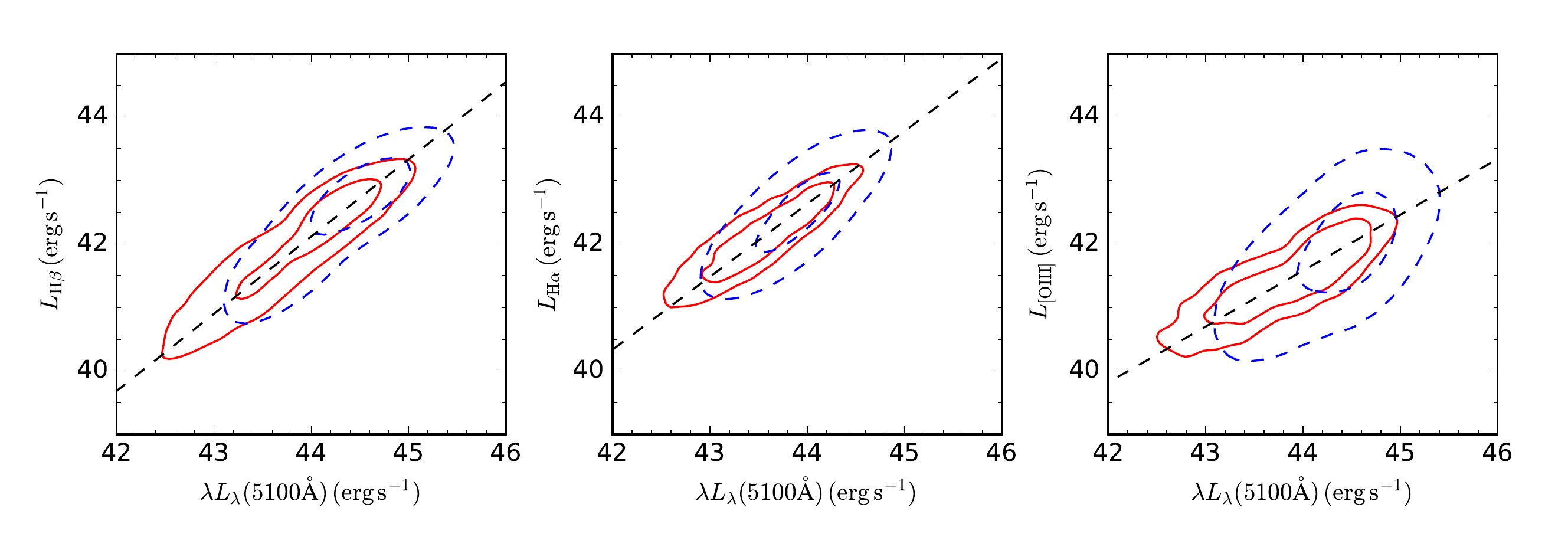}
\caption{From left to right, the luminosity of H$\beta$, H$\alpha$ and [O III] ($\lambda 5007$\AA) plotted against the monochromatic luminosity at 5100 \AA \, 
for NLSy1 galaxies (solid red contours) and BLSy1 galaxies (dashed blue contours). The dashed black line is the linear fit to the data considering all objects. The 68 percentile (inner) and 95 percentile (outer) density contours are shown in each panel.}
\label{fig_2}
\end{figure}

In Figure \ref{fig_2}, we have plotted the luminosity of several emission lines against the monochromatic nuclear continuum luminosity ($\lambda L_{5100}$). Luminosity of H$\beta$ (left), H$\alpha$ (middle) and [O III] (right) is plotted against $\lambda L_{5100}$. All the correlations are very strong which remain valid even if we divide the entire sample into smaller redshift bins. Linear least squares fits yield
\begin{eqnarray}
\log(L_{\mathrm{H\beta}})=( 1.217 \pm 0.003 ) \times \log(\lambda L_{5100}) +( -11.43 \pm 0.13)\\
\log(L_{\mathrm{H\alpha}})=( 1.148 \pm 0.006 ) \times \log(\lambda L_{5100}) +( -7.87 \pm 0.26)\\
\log(L_{\mathrm{[O III]}})=( 0.880 \pm 0.005 ) \times \log(\lambda L_{5100}) +( 2.84  \pm 0.22)
\end{eqnarray}

These results are consistent with the relation found by June et al. (2015).
According to June et al. (2015), the relation between line and continuum 
luminosity is valid over a wide luminosity and redshift ($z=0-6$) ranges. This means 
that the response of the gas clouds in the broad line region (BLR) to the 
incident ionizing continuum is the same over a wide redshift range. Such strong
 correlations between the line and the continuum luminosity have a significant 
implication on $M_{\mathrm{BH}}$ measurements in AGN using the 
virial relation. In some AGNs where the host galaxy contamination is large and the emission from the relativistic jet is significant, estimation of the 
continuum luminosity would be difficult and thus the emission line luminosity 
could be used as a surrogate for the continuum luminosity. Hence, the correlation 
between the luminosity of [O III] and $\lambda L_{5100}$ is crucial as it can 
be used to estimate the bolometric luminosity and $M_{\mathrm{BH}}$ in Seyfert 2 AGNs. The $M_{\mathrm{BH}}$ values for all objects were estimated using the virial 
relationship given as 
\begin{equation}
M_{\mathrm{BH}}=f\frac{R_{\mathrm{BLR}} \Delta v^2}{G}
\end{equation}
where $f$ is an unknown geometrical factor that depends on 
the BLR geometry and kinematics (see Rakshit et al. 2015 and the references therein), $\Delta v$ is the 
FWHM of the broad emission line and $R_{\mathrm{BLR}}$ is the radius of the 
BLR estimated using the reverberation mapping scaling 
relation (Bentz et al. 2013)
\begin{equation}
\log R_{\mathrm{BLR}} (\mathrm{lt-day}) = 1.527 + 0.533 \times \log \left(\frac{\lambda L_{\lambda} (5100 \AA)}{10^{44}} \mathrm{erg\, s^{-1}}     \right).
\end{equation}
We used $f=0.75$ considering a spherical distribution of clouds and estimated $M_{\mathrm{BH}}$. We also calculated the Eddington ratio ($\xi_{\mathrm{Edd}}$), which is 
defined as $\xi_{\mathrm{Edd}}=L_{\mathrm{bol}}/L_{\mathrm{Edd}}$ where 
$L_{\mathrm{bol}}= 9\times \lambda L_{\lambda} (5100)$ (Kaspi et al. 2000) 
and $L_{\mathrm{Edd}}=1.3 \times 10^{38} M_{\mathrm{BH}}/M_{\odot}$ erg $\mathrm{s}^{-1}$.

\begin{figure}[h]
\centering
\includegraphics[width=12cm, height=5cm]{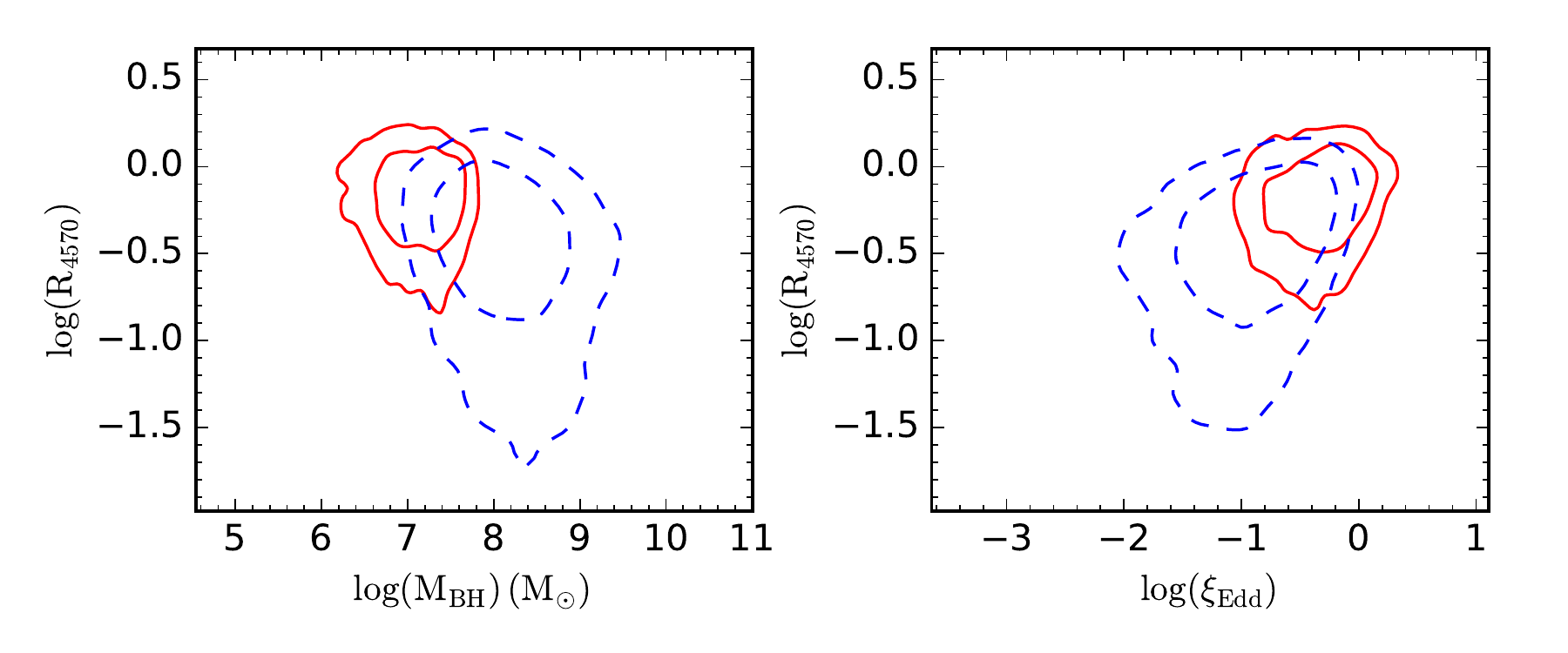}
\caption{The relation between $R_{4570}$ and black hole mass (left) and Eddington ratio (right) for NLSy1 galaxies (solid red contours) and BLSy1 galaxies (dashed blue contours). The 68 percentile (inner) and 95 percentile (outer) density contours are shown in each panel.}\label{fig_3}
\end{figure}

We show in Figure 3, the variation of $R_{4570}$ against $M_{\mathrm{BH}}$ (left panel) 
and the Eddington ratio (right panel) for both NLSy1 and BLSy1 galaxies. The 
68 percentile (inner) and 95 percentile (outer) density contours are also 
shown in each panel. It is clear from Figure 3 that NLSy1 galaxies occupy a unique 
location in the $R_{4570}-M_{\mathrm{BH}}$ and $R_{4570}-\xi_{\mathrm{Edd}}$ 
plane. They have a lower black hole mass, higher Fe II strength and  higher 
Eddington ratio compared to their broad line counterparts. Thus,  in the $R_{4570}-M_{\mathrm{BH}}$ diagram, NLSy1 galaxies are situated at the extreme top left while in the $R_{4570}-\xi_{\mathrm{Edd}}$ diagram their location is at the top right corner. Thus, one can use the $R_{4570}-M_{\mathrm{BH}}$ and 
$R_{4570}-\xi_{\mathrm{Edd}}$ diagrams to distinguish between
NLSy1 and BLSy1 galaxies.

\section{Radio properties}
\begin{figure}[h]
\centering
\includegraphics[width=12cm, height=5.0cm]{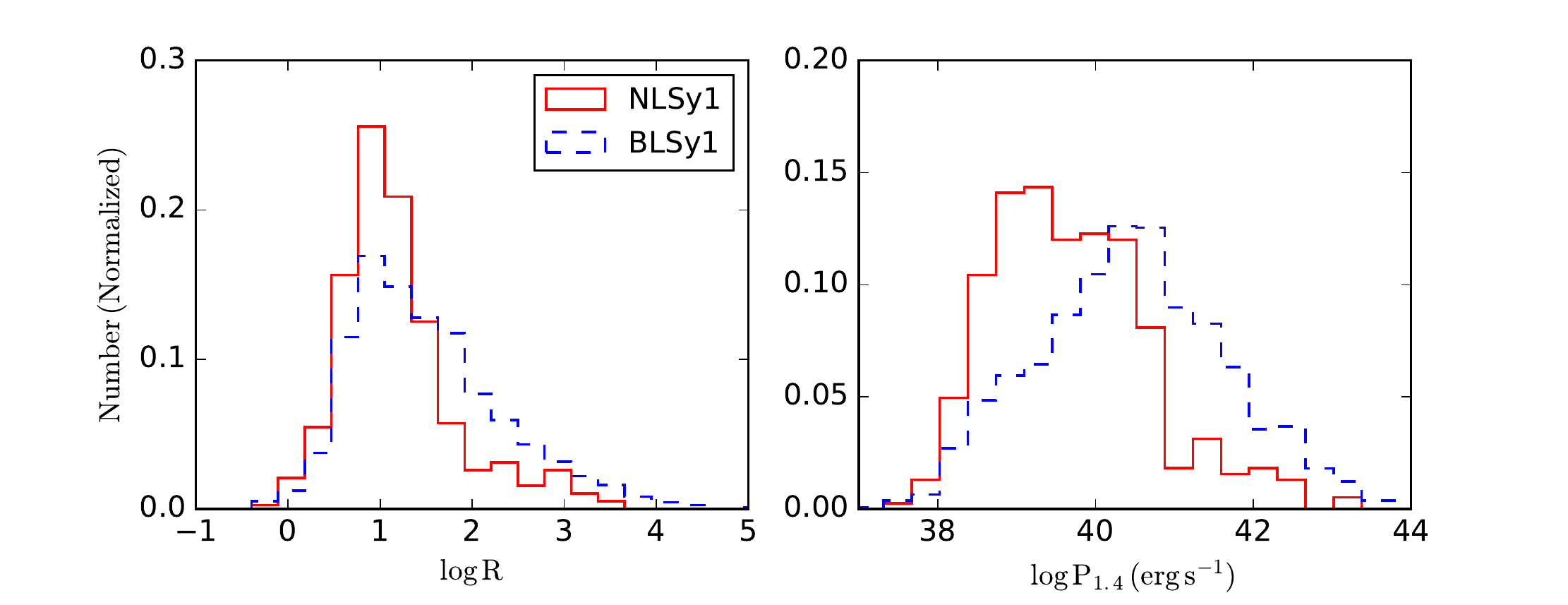}
\caption{The distribution of radio loudness (left) and radio power (right) for NLSy1 (solid-line) and BLSy1 (dashed-line) galaxies.}\label{fig:radio}
\end{figure}

BLSy1 galaxies usually show stronger radio emission than NLSy1 galaxies (Komossa
et al. 2006). In Rakshit et al. (2017a) we found that about 5\% of the NLSy1 galaxies have been detected in the FIRST survey within a search radius of 2$^{{\prime}{\prime}}$. 
Among the sources studied in this work, 383 NLSy1 galaxies and 1547 BLSy1 galaxies have been detected by FIRST. We calculated their radio loudness parameter ($R$) using $R=F_{1.4 \mathrm{GHz}}/F_g$, where $F_{1.4 \mathrm{GHz}}$ is the 1.4GHz radio flux density and $F_g$ is the flux density in the  optical g-band. 
Their radio loudness distribution is plotted in the left panel of Figure \ref{fig:radio}. We found that the $R$ distribution in the case of NLSy1 drops much rapidly after $R\sim10$ compared to BLSy1 galaxies. Both the distributions however peak at $R\sim10$. The radio power ($P_{1.4 \mathrm{GHz}}$) was calculated following Rakshit et al. (2017a) and plotted in the right panel of Figure \ref{fig:radio}. The histograms clearly show that BLSy1 galaxies have powerful jets compared to  NLSy1 galaxies. 

\section{X-ray properties}
\begin{figure}[h]
\centering
\includegraphics[width=17cm, height=4.2cm]{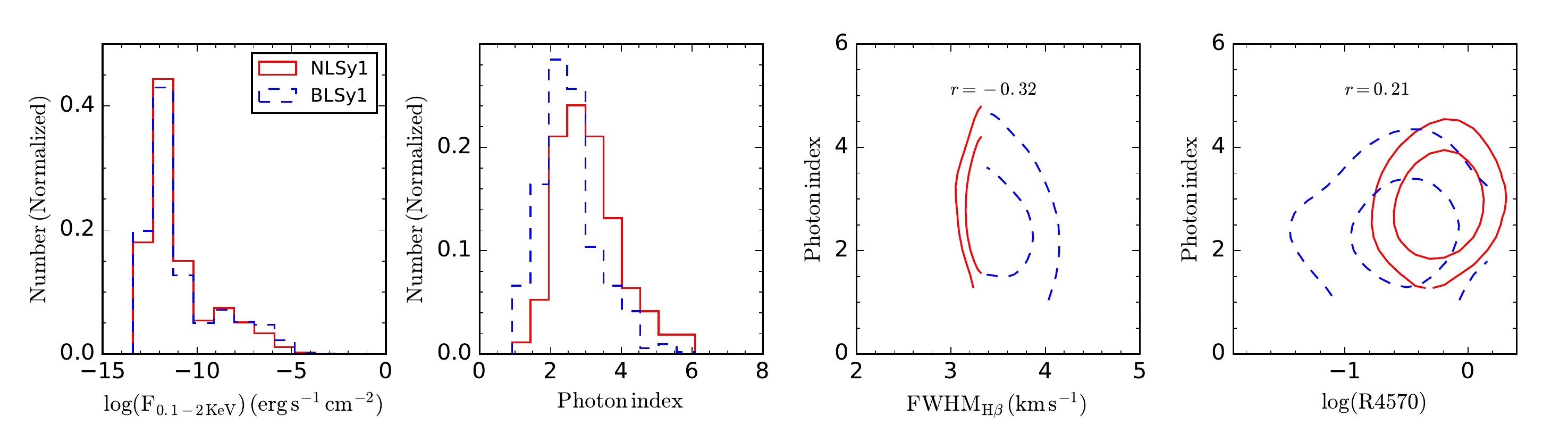}
\caption{From left to right the distribution of the soft X-ray (0.1$-$2 keV) 
flux, the distribution of the photon index, the variation of the photon index with 
the FWHM of the H$\beta$ broad component, and the variation of the photon index 
with $R_{4570}$ is plotted. The Spearman rank correlation coefficients are given
in each panel. The 68 percentile (inner) and 95 percentile (outer) density contours are shown in the third and fourth panels.}\label{fig:xray}
\end{figure}

One of the important characteristics of NLSy1 galaxies is that they exhibit 
rapid X-ray variability and have soft X-ray spectra (see Boller et al. 1996; Leighly 1999). To study the X-ray properties of the objects in our sample we 
looked for their X-ray counterparts in the ROSAT all-sky survey (2RXS) source catalog (Boller et al. 2016) within a search radius of 30$^{{\prime}{\prime}}$ 
(see Rakshit et al. 2017a for more details). We found that about 1300 NLSy1 and 3600 BLSy1 galaxies 
in our sample studied here have X-ray counterparts in ROSAT. We plotted their 
soft X-ray (0.1$-$2 keV) flux in the first panel of Figure \ref{fig:xray}. Both the distribution of soft X-ray flux for NLSy1 and BLSy1 galaxies are similar. 
The distribution of their photon indices taken from the 2RXS catalog 
is shown in the second panel. The photon index distribution has a median 
value of $\Gamma$ = $2.9\pm0.9$ for NLSy1 galaxies and 
$\Gamma$ = $2.4\pm0.8$ for BLSy1 galaxies suggesting that 
NLSy1 galaxies on average have a steeper soft X-ray spectrum than BLSy1 galaxies.
This result from an analysis of a large sample of objects is consistent with the findings of Leighly (1999). In the third panel of Figure \ref{fig:xray}, 
$\Gamma$ is plotted against the width of the H$\beta$ broad component for NLSy1 and BLSy1 galaxies. We found that $\Gamma$ is anti-correlated with 
the H$\beta$ width suggesting that the NLSy1 galaxies have $\Gamma$ larger 
than BLSy1 galaxies. The Spearman rank correlation gives a value of $-0.32$ 
suggesting a moderate anti-correlation between the two parameters. In the last 
panel, $\Gamma$ is plotted against $R_{4570}$. We find a weaker but positive correlation between the two having a Spearman rank correlation coefficient of 
0.21 suggesting that strong Fe II emitters have a higher $\Gamma$, which is 
larger for NLSy1 galaxies than for  BLSy1 galaxies.      

\section{Optical variability properties}
AGNs are known to show rapid variability across all wavelengths over  time 
scales of days to years. However, the origin of such variations is poorly 
understood. To understand this, several studies have been conducted using a 
large sample of Seyfert 1 galaxies. The amplitude of the optical variability 
is found to be correlated with several observables such as wavelength, 
luminosity, redshift, $M_{\mathrm{BH}}$ and Eddington ratio (see MacLeod et al. 2010 
and references therein). However, studies on the optical variability of 
NLSy1 galaxies are very limited. 
Klimek et al. (2004) and more 
recently Ai et al. (2010,2013) have performed an optical variability study of 
a small sample of NLSy1 galaxies and found that they are less variable than
 BLSy1 galaxies. As NLSy1 galaxies have a low $M_{\mathrm{BH}}$ and a high 
Eddington ratio, variability studies of NLSy1 galaxies along with
BLSy1 galaxies will enable one to probe the variability characteristics 
of AGNs over a wide range of $M_{\mathrm{BH}}$ and Eddington ratio.

\begin{figure}[h]
\centering
\includegraphics[width=16cm, height=5.5cm]{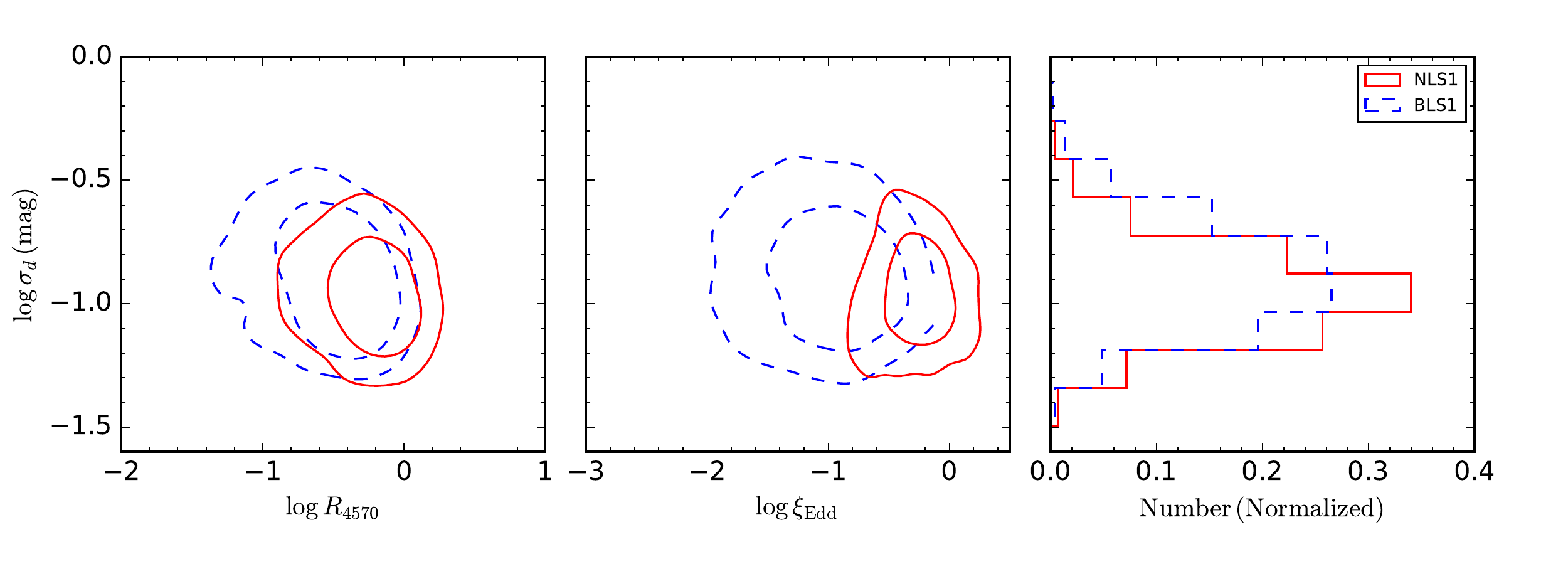}
\caption{Density contours of the amplitude of variability against $R_{4570}$ (left) and $\xi_{\mathrm{Edd}}$ (middle) for 68 (inner) and 95 percentile (outer) of the sample. The right panel shows the distribution of the amplitude of variability for NLSy1 and BLSy1 galaxies. Figure adapted from Rakshit et al. (2017b).}\label{fig:optical}
\end{figure}

Using the extended NLSy1 galaxies catalog of Rakshit et al. (2017a), we have 
recently performed a comparative study of the optical variability of a large 
sample of NLSy1 and BLSy1 galaxies matched in luminosity and redshift 
(Rakshit et al. 2017b). For this purpose, we used the optical V-band light 
curves from the Catalina Real Time Transient Survey (CRTS) covering more than 
5 years of observation and having a minimum of 50 photometric data points in each 
light curve. The light curves were modeled using the JAVELIN code which uses a 
damped random walk model (see Zu et al., 2011 and references therein) allowing 
to estimate the amplitude of variability ($\sigma_d$) of all the BLSy1 and NLSy1 galaxies. In addition, the intrinsic amplitude of variability ($\sigma_m$) was estimated from the measured variance of the observed light curves after subtracting the measurement errors. The $\sigma_d$ calculated using JAVELIN is found to be consistent with the value of $\sigma_m$ (see Rakshit et al., 2017b). The optical variability amplitude is found to be lower in NLSy1 galaxies compared 
to BLSy1 galaxies.
In Figure \ref{fig:optical}, we show the dependency of the amplitude of 
variability with $R_{4570}$ and $\xi_{\mathrm{Edd}}$ in the left and middle 
panels. The amplitude of variability is anti-correlated with both $R_{4570}$ 
and $\xi_{\mathrm{Edd}}$. The right panel shows the distribution of amplitude 
of variability for NLSy1 and BLSy1 galaxies clearly stating that on average NLSy1 
galaxies are less variable in the optical band than BLSy1 galaxies. 
As $R_{4570}$ and the Eddington ratio are strongly correlated it is likely that
the difference in the variability between NLSy1 and BLSy1 galaxies might be
in part due to differences in the Eddington ratio between them. To investigate
this, we created a subsample of NLSy1 and BLSy1 galaxies matched in 
Eddington ratio. In this subsample, we find no difference in variability between BLSy1 and
NLSy1 galaxies, thereby, suggesting the Eddington ratio as the most fundamental
parameter driving optical variabiltiy. 
More details can be found in Rakshit et al. (2017b).              
 
\section{Conclusion}
We have studied the emission line and variability properties of a large sample 
of NLSy1 and BLSy1 galaxies from SDSS-DR12. The emission line parameters 
were taken from the work of Rakshit et al. (2017a). Our sample
consists of  4070 NLSy1 and 14,314 BLSy1 galaxies having a median S/N $>10 
\, \mathrm{pixel^{-1}}$ and $R_{4570}>0.01$. The results of this work are
summarized below. 
\begin{enumerate}
\item The line widths of H$\beta$ and H$\alpha$ are strongly correlated 
via the relation $\mathrm{FWHM(H\alpha)}= (0.768$\\$ \pm 0.004) 
\times \mathrm{FWHM(H\beta)}$ for the entire sample. Also, a strong 
correlation is found between the fluxes of H$\alpha$ and H$\beta$ lines 
in both NLSy1 and BLSy1 galaxies. For the combined sample 
we find $\mathrm{flux(H\alpha)=(3.09 \pm 0.01) \times flux(H\beta)}$.

\item The luminosity of the Balmer lines as well as [O III] is found to be 
strongly correlated with the continuum luminosity at 5100 \AA. This suggests 
that the response of the BLR on the ionizing continuum is 
identical over a wide range of redshift and luminosity.  

\item NLSy1 galaxies have a higher Fe II strength, lower $M_{\mathrm{BH}}$ and 
higher Eddington ratio compared to BLSy1 galaxies. These characteristics 
place them at the extreme top left corner in the $R_{4570}-M_{\mathrm{BH}}$ and top right corner in the $R_{4570}-\xi_{\mathrm{Edd}}$ diagrams.

\item The radio loudness distribution of both the NLSy1 and BLSy1 galaxies 
peaks at about R = 10 but drops rapidly after $R\sim10$ for NLSy1 compared to BLSy1 galaxies. The latter show more powerful jets compared to the former.

\item NLSy1 galaxies on average have a higher photon index ($\Gamma$ = 
$2.9\pm0.9$) or steeper soft X-ray spectrum compared to the BLSy1 galaxies 
which have a $\Gamma$ of $2.4\pm0.8$. A moderate anti-correlation between 
$\Gamma$ and the width of the emission line is found considering the whole 
sample. The photon index is positively correlated with the $R_{4570}$. 

\item NLSy1 galaxies on average have a lower amplitude of variability compared to the BLSy1 galaxies and it is anti-correlated with the $R_{4570}$ and 
$\xi_{\mathrm{Edd}}$. Our analysis indicates that the $\xi_{\mathrm{Edd}}$ plays an
important role in driving optical flux variations in AGNs.
\end{enumerate}         

\section*{Acknowledgements}
We are grateful to the anonymous referee and Jean Surdej (guest co-editor) for their suggestions on our manuscript. We are thankful to the organizers of ``The First BINA Workshop'' in ARIES, Nainital, India for providing an opportunity to present our work. S.R. thanks Neha Sharma for carefully reading the manuscript. 

%
%
%
%
%

\footnotesize
\beginrefer
\refer Osterbrock, D. E., \& Pogge, R. W. 1985, ApJ, 297, 166

\refer Boller, T., Brandt, W. N., \& Fink, H. 1996, A\&A, 305, 53

\refer Leighly, K. M. 1999, ApJS, 125, 317

\refer Zhou, H., Wang, T., Yuan, W., et al. 2006, ApJS, 166, 128

\refer Xu, D., Komossa, S., Zhou, H., et al. 2012, AJ, 143, 83

\refer Rakshit, S., Stalin, C. S., Chand, H., et al. 2017, ApJS, 229, 2

\refer Baldi, R. D., Capetti, A., Robinson, A., et al. 2016, MNRAS, 458, 69

\refer Turner, T. J., George, I. M., Nandra, K., et al. 1999, ApJ, 524, 667

\refer Grupe, D., Beuermann, K., Mannheim, K., \& Thomas, H.-C. 1999, A\&A, 350, 805


\refer Bruzual, G., \& Charlot, S. 2003, MNRAS, 344, 1000

\refer Kovacevic, J., Popovic, L. C., \& Dimitrijevic, M. S. 2010, ApJS, 189, 15

\refer Jun, H. D., Im, M., Lee, H. M., et al. 2015, ApJ, 806, 109

\refer Kauffmann, G., Heckman, Timothy M., Tremonti, C., et al., 2003, MNRAS, 346, 1055

\refer Rakshit, S., Petrov, R. G., Meilland, A., \& Honig, S. F. 2015, MNRAS, 447, 2420

\refer Bentz, M. C., Denney, K. D., Grier, C. J., et al. 2013, ApJ, 767,
149

\refer Kaspi, S., Smith, P. S., Netzer, H., et al. 2000, ApJ, 533, 631

\refer Komossa, S., Voges, W., Xu, D., et al. 2006, AJ, 132, 531

\refer Boller, T., Freyberg, M. J., Trumper J., et al. 2016, A\&A, 588, 103

\refer MacLeod, C. L., Ivezic, Z., Kochanek, C. S., et al. 2010, ApJ, 721, 1014

\refer Klimek, E. S., Gaskell, C. M., \& Hedrick, C. H. 2004, ApJ, 609, 69

\refer Ai, Y. L., Yuan, W., Zhou, H. Y., et al. 2010, ApJL, 716, L31

\refer Ai, Y. L., Yuan, W., Zhou, H., et al. 2013, AJ, 145, 90

\refer Rakshit, S. and Stalin, C. S., 2017, ApJ, accepted

\refer Zu, Y., Kochanek, C. S., \& Peterson, B. M. 2011, ApJ, 735, 80

\endrefer           

\end{document}